\documentclass[twocolumn,aps,showpacs,amsmath,amssymb,preprintnumbers]{revtex4}

\usepackage{graphicx}
\usepackage{dcolumn}
\usepackage{bm}

\def\ep{\epsilon}

\def\order#1{{\cal O}\!\left(#1\right)}

\newcommand{\ba}{\begin{eqnarray}}
\newcommand{\ea}{\end{eqnarray}}
\newcommand{\beq}{\begin{equation}}
\newcommand{\eeq}{\end{equation}}

\begin{document}
\preprint{Alberta Thy 12-08}

\title{Mass effects in  muon and semileptonic \boldmath $b\to c$
  decays}

\author{Alexey Pak}
\author{Andrzej Czarnecki}
\affiliation{Department of Physics, University of Alberta\\ 
  Edmonton, AB\ T6G 2G7, Canada}
 
\begin{abstract}
Quantum chromodynamics (QCD) effects in the semileptonic decay 
$b\to c \ell \bar{\nu}$ are evaluated to the second order in the coupling 
constant, $\order{\alpha_s^2}$, and to several orders in the expansion 
in quark masses, $m_c/m_b$.  Corrections are calculated for the 
total decay rate as well as for the first two moments of the lepton 
energy and the hadron system energy distributions.  Applied to the muon
decay, they decrease its predicted rate by $-0.43$ ppm.  
\end{abstract}

\pacs{13.35.Bv,13.25.Hw,12.38.Bx}

\maketitle

Decays of heavy fermions are an abundant source of information about
fundamental interactions.  Particularly important among them is the
muon ($\mu$) decay.  Insensitive to strong interactions, it can be
very precisely described by the electroweak model. The experiment
MuLan at the Paul Scherrer Institute will likely measure the rate of
the muon decay with an uncertainty better than 1 ppm and thus improve
the determination of the Fermi constant $G_F$ that describes the
strength of the charged-current weak interaction
\cite{Chitwood:2007pa}.  Along with the fine structure constant
$\alpha$ and the $Z$-boson mass, $G_F$ is one of the three pillars of
electroweak Standard Model tests \cite{Marciano:1999ih}.

In a separate effort, the TWIST experiment at TRIUMF measures energy
and angular distributions of positrons in the $\mu^+$ decay,
testing the Standard Model and searching for new
interactions, notably new bosons predicted by left-right symmetric
models \cite{Musser:2004zw,Gaponenko:2004mi,Jamieson:2006cf}.

To match this experimental progress, both the rate
\cite{vanRitbergen:1998yd} and the energy distribution
\cite{Anastasiou:2005pn} have been calculated in quantum
electrodynamics (QED) with $\order{\alpha^2}$
accuracy.  Two-loop weak corrections have also been calculated
\cite{Awramik:2003ee}. In the decay rate studies,  the
electron mass $m_e$  was assumed negligible in the already small
$\order{\alpha^2}$ effects.

Here we show that the finite $m_e$ effect decreases the muon
decay rate by about half ppm,  exceeding previous 
estimates \cite{vanRitbergen:1999fi} and approaching 
the expected MuLan precision.

The final-state fermion mass effects are much larger in the
heavy-quark decay $b\to c \ell \bar{\nu}$.  Studied in $B$-factories
and the Tevatron, this process provides information about the
Cabibbo-Kobayashi-Maskawa (CKM) matrix element $V_{cb}$, as well as
about parameters governing heavy-quark dynamics (see
\cite{Neubert:2008cp} for an up to date review and references).  Also
in this case, theoretical studies at $\order{\alpha_s^2}$ are complete
only for a massless final-state quark \cite{vanRitbergen:1999gs}.  For
the actual massive $c$-quark, $\order{\alpha_s^2}$ effects are known
in some special cases of kinematics
\cite{Czarnecki:1996gu,Czarnecki:1997hc,Czarnecki:1998kt}.  So-called
Brodsky-Lepage-Mackenzie (BLM) corrections \cite{Brodsky:1982gc} have
been obtained for the width \cite{Luke:1994du} and moments of the
energy spectrum \cite{Falk:1997jq,Gambino:2004qm}.  Also some
logarithms of the mass $m_c$ have been determined to all orders in
$\alpha_s$ \cite{Bauer:1996ma}.  Most recently, Melnikov calculated
numerically the $m_c$ effects for the width and the first two moments
of the energy distribution of hadrons and of the charged lepton
produced in this decay \cite{Kirill2008}.  In this paper we present
corresponding analytical results obtained as an expansion in powers
and logarithms of $\rho\equiv m_c/m_b$.

\begin{figure}[t]
  \parbox[t]{0.24\textwidth}{
    \includegraphics[width=0.17\textwidth]{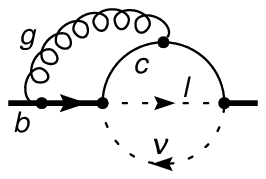}} 
  \put(-20,45){\makebox(0,0){(a)}}
  \parbox[t]{0.24\textwidth}{
    \includegraphics[width=0.17\textwidth]{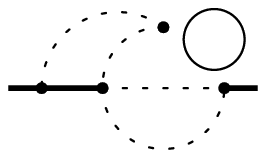}} 
  \put(-100,45){\makebox(0,0){(b)}} \\
  \parbox[t]{0.24\textwidth}{
    \includegraphics[width=0.17\textwidth]{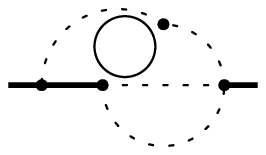}} 
  \put(-20,45){\makebox(0,0){(c)}}
  \parbox[t]{0.24\textwidth}{
    \includegraphics[width=0.17\textwidth]{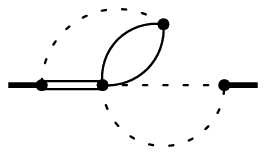}}
  \put(-100,45){\makebox(0,0){(d)}} \\
  \parbox[t]{0.48\textwidth}{\caption{\label{fig:regions}
  Example of an expansion of a double-scale integral.
  Thick, thin, and dashed lines correspond to $m_b$, $m_c$, and 
  massless propagators. 
Loop momenta can be all hard  (Taylor expansion in $m_c$ of (a)), 
  and one or more soft ((b), (c), and (d)). Double line in
  (d) denotes a static propagator.}}
\end{figure}
The construction of such mass expansion is illustrated with an
$\order{\alpha_s}$ example in
Fig.~\ref{fig:regions}.   Real and virtual corrections are calculated
together as cuts of the diagram \ref{fig:regions}(a).
Depending on virtualities of momenta flowing into and through the
charm quark lines, $m_c$ can be treated as small compared to those
momenta, or else those momenta can be treated as small compared to
the $b$-quark mass in other lines of the diagram.  The most
interesting case is shown in Fig.~\ref{fig:regions}(d), where two loop
momenta are of order $m_c$.  This configuration generates odd powers
of $\rho$ and will be discussed in some detail below.

\begin{figure}[t!]
\begin{center}
  \includegraphics[width=0.19\textwidth]{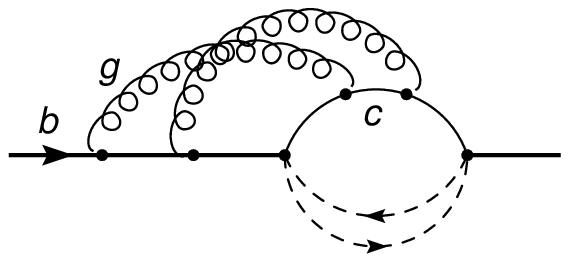}\hspace{0.2cm}
  \includegraphics[width=0.19\textwidth]{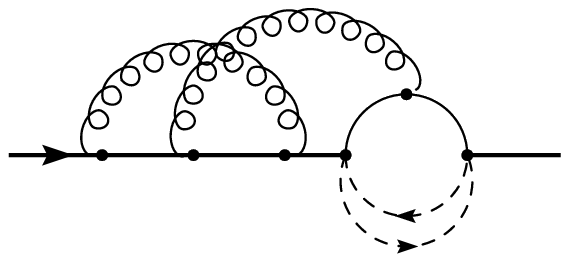}\\
  \includegraphics[width=0.19\textwidth]{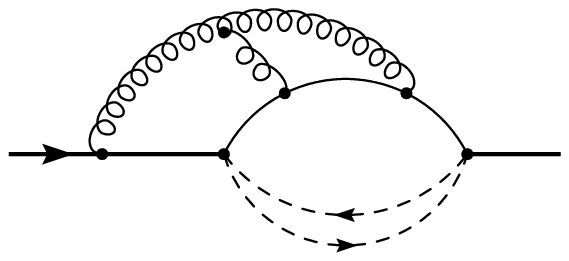}\hspace{0.2cm}
  \includegraphics[width=0.19\textwidth]{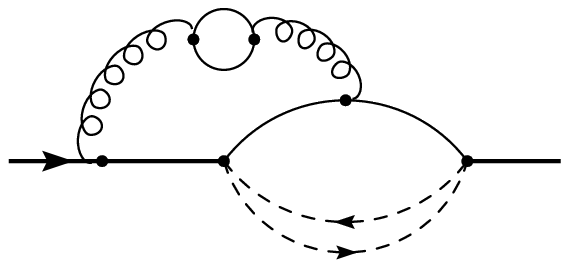}\\
  \parbox[t]{0.48\textwidth}{\caption{\label{fig:diags}
    Examples of diagrams contributing to the 
    $\mathcal{O}(\alpha_s^2)$ corrections to the $b$-quark decay rate.}}
\end{center}
\end{figure}

At $\mathcal{O}(\alpha_s^2)$ the number of diagrams is larger and the
analysis of momentum scales more challenging, with as many as 11
regions in some diagrams, but it follows the general pattern outlined
above.  As a result, even the four-loop diagrams shown in 
Fig.~\ref{fig:diags} are calculated as a series in $\rho$ with exact
coefficients.   

The most challenging contributions arise when all loop momenta are hard
($\sim m_b$).  All charm-quark lines are Taylor-expanded in $m_c$,
generating a large number of integrals with various powers 
of denominator factors.  The method of 
Ref.~\cite{Laporta:2001dd} is employed to reduce all integrals to a
set of master integrals. Most are known
\cite{vanRitbergen:1999fi}, sufficing  to reproduce the 
$m_c = 0$ limit as well as the  $\mathcal{O}(\rho^2)$ 
terms. Unfortunately, the following 
term $\mathcal{O}(\rho^4)$ 
requires further $\ep$-expansion of the most difficult integrals
(calculations are conducted in  $D=4-2\ep$ dimensions). 

An approach to evaluating the additional terms is illustrated with the
the first diagram in Fig.~\ref{fig:diags}, normalized at $p^2=-1$.  
After integrating the lepton loop momentum,
\begin{eqnarray}\label{eqn:t3be56}
&& I(a_1,a_2,a_3,a_4,a_5,a_6,a_7,a_8,a_9;x) \nonumber \\
&=& \int\frac{d^D k_1 d^D k_2 d^D k_3 ~ D_9^{- a_9}}{
  D_1^{a_1 + \ep}D_2^{a_2}D_3^{a_3}D_4^{a_4}
  D_5^{a_5}D_6^{a_6}D_7^{a_7}D_8^{a_8}},
\end{eqnarray}
with $D_1 = k_1^2$, $D_2 = (k_1 + p)^2$, $D_3 = (k_1 + k_2 + p)^2$,
$D_4 = (k_1 + k_2 + k_3 + p)^2$, $D_5 = (k_2 + k_3 + p)^2 + x$,
$D_6 = (k_3 + p)^2 + x$, $D_7 = k_2^2$, $D_8 = k_3^2$, and $D_9 =
2 k_2 p$.  The artificial scale $x$, introduced in some
propagators, must be set equal one for   the needed on-shell case.

This double-scale topology has 40 master integrals, instead of 21 in
the single-scale case.  40 differential equations are set up by
relating master integrals to their derivatives, using $-\partial_x
I(\{a_i\};x) = (a_5 {\bf 5^+} + a_6{\bf 6^+})I(\{a_i\}; x)$ (operator
${\bf i^+}$ increments exponent $a_i$).  Fortunately, this
system splits into independent subsystems of at most four equations.
Results are expanded
to the needed order in $\ep$ \cite{Remiddi:1997ny}, and 
integration constants are fixed using
the behavior near $x \to \infty$ \cite{vanRitbergen:1999fi}. Finally, the
limit $x \to 1$ is taken.

In addition to this all-hard case, there are regions with some hard
and some soft momenta, factorizing into diagrams with less than
four loops.  Among them the most difficult ones have three soft loop
momenta and are evaluated using
Ref.~\cite{Grozin:2006xm,Grozin:2007ap,SmirnovPrivate}.

Up to the two-gluon order, QCD corrections to the process $b\to
c\ell\bar{\nu}$ are parameterized by 
\begin{equation*}
  \Gamma(b\to c\ell\bar{\nu}) = \Gamma_0 C_F\left[{X_0\over C_F} +
  \frac{\alpha_s}{\pi} X_1 +
  \left(\frac{\alpha_s}{\pi}\right)^2 X_2 + \ldots
\right],
\end{equation*}
where $\Gamma_0 = \frac{G_F^2 |V_{cb}|^2 m_b^5}{192\pi^3}$ is the
tree-level massless result, and $\alpha_s$ is normalized in the 
$\overline{\rm MS}$ scheme at $m_b$. The tree-level mass-dependent 
decay rate is 
\begin{equation}\label{eq:tree}
X_0 = 1 - 8\rho^2 - 24\rho^4\ln\rho + 8\rho^6 - \rho^8.
\end{equation}
The one-gluon correction, known exactly \cite{Nir:1989rm}, has an
expansion in $\rho$ starting with
\begin{eqnarray}\label{eq:X1}
\lefteqn{ X_1 = \frac{25}{8} - \frac{\pi^2}{2} 
    - \left(34 + 24\ln\rho\right)\rho^2 
    + 16\pi^2\rho^3 }
  \nonumber \\  &&  
    - \left(\frac{273}{2} - 36\ln\rho + 72\ln^2\rho + 8\pi^2 \right)\rho^4
    + \ldots
\end{eqnarray}
The second order correction $X_2$ is a sum of finite, gauge invariant 
parts proportional to various color factors,
\begin{eqnarray}\label{eqn:X2}
   X_2 = C_F X_A +  C_A X_N +  T_R\left( 
      X_C + X_H + N_L X_L \right).
\end{eqnarray}
$X_L$, $X_C$, and $X_H$ denote contributions of $c$-, $b$-, and $N_L =
3$ species of massless quarks. SU(3) color factors are $T_R =
\frac{1}{2}$, $C_F = \frac{4}{3}$, $C_A = 3$.

Using techniques described above, 
expansions of $X_A$, $X_N$, $X_C$, $X_L$, and $X_H$ are obtained
through $\mathcal{O}(\rho^7)$, sufficient for sub-percent
accuracy of $X_2$ for the physical $\rho \approx 0.3$. These
expressions being rather lengthy, Table~\ref{tab:results} lists
numerical values of 
the coefficients.  $X_{N}$ and $X_C$
are shown explicitly for
the purpose of subsequent discussion,
\begin{eqnarray}\label{eqn:xn}
  X_C &=&
    - \frac{1009}{288} + \frac{8}{3}\zeta_3 + \frac{77}{216}\pi^2
    - \frac{5}{4}\pi^2\rho
    \nonumber \\ 
    &+& \left[\frac{145}{3} + \frac{52}{3}\ln\rho
        - 8\ln^2\rho + \frac{16}{3}\pi^2 \right] \rho^2
    \nonumber \\
    &+& \left[{569\over 36} + {64\over 3}\ln\rho\right]\pi^2\rho^3 
    + \ldots,
\nonumber \\
   X_N &=& \frac{154927}{10368} - \frac{383\zeta_3}{72} + \frac{95\pi^2}{162} 
      - \frac{53\pi^2\ln{2}}{12} + \frac{101\pi^4}{1440} 
  \nonumber \\
    &+& \left[
      \frac{539\pi^4}{1080} 
      - \frac{1181\pi^2}{216} 
      - \frac{185}{3}\ln\rho 
      + 22\ln^2\rho  
      - 43\zeta_3 
  \right. \nonumber \\ 
      &+& \left. 4\pi^2\ln{2} 
      - \frac{1537}{16}  
      \right]\rho^2 
    + \left[
      \frac{556\pi^2}{3} 
      - \frac{1136\pi^2\ln{2}}{3} 
  \right. \nonumber \\
      &+& \left. \frac{56\pi^3}{3}
      - \frac{124\pi^2}{3}\ln\rho 
      \right]\rho^3 
    + \left[
      \frac{1777\pi^4}{720} 
      - \frac{23807}{864}
  \right. \nonumber \\ 
     &+& \left(
	\frac{577}{36} 
	- 39\zeta_3 
	- \frac{10\pi^4}{3} 
	+ 30\pi^2\ln{2} 
	- 15\pi^2
	\right)\ln\rho
  \nonumber \\
      &+& \left(5\pi^2 - 185\right)\ln^2\rho
      + 88\ln^3\rho 
      + 4\pi^2\ln^2{2}
  \nonumber \\ 
      &+& 48\mbox{Li}_4\frac{1}{2} 
      - \frac{215\pi^2\zeta_3}{6} 
      + \frac{727\pi^2}{48} 
      - \frac{535\zeta_5}{2} 
      - \frac{615\zeta_3}{4} 
  \nonumber \\
      &+& \left. 2\ln^4 2 
      - \frac{13\pi^2\ln{2}}{2} 
    \right]\rho^4
    + \dots~.
\end{eqnarray}
These correction significantly depend on $\rho$ near its 
realistic values: $X_2(0.25) = -6.59$, while $X_2(0.3) = -4.89$.

In addition to the decay rate, corrections to the first two moments in
lepton energy $\hat{E}_l = E_l/m_b$, and the
hadronic system energy $\hat{E}_h = E_h/m_b$, have been computed.  
The average is taken over the whole phase space of
decay products. These moments are parameterized by
\begin{eqnarray*}
{1\over \Gamma_0} \langle \hat{E}_l^n
\rangle
\equiv
\sum_{j=0}^\infty \left(\alpha_s\over\pi\right)^j L_j^{(n)},
\end{eqnarray*}
and similarly for the moments of $\hat{E}_h$, described by coefficients
$H^{(n)}_j$. 
Table~\ref{tab:results} shows 
the second-order corrections $L^{(1,2)}_2$ and $H^{(1,2)}_2$.

Finally, Table~\ref{tab:results} lists $U_C$, the $c$-quark contribution
to $\Gamma(b\to u\ell\bar{\nu})$, defined by analogy with $X_C$ 
of (\ref{eqn:X2}). This result is useful e.g. in $b\to s\gamma$ 
studies \cite{Misiak2007}.

\begingroup
\squeezetable
\begin{table*}[tbh]
\begin{ruledtabular}
\begin{tabular}{l  r  r  r  r  r  r  r  r  r  r  r}
& $\rho^0$ & $\rho^1$ & $\rho^2$ & $\rho^2\ln\rho$ & $\rho^2\ln^2\rho$ & $\rho^3$ & 
$\rho^3\ln\rho$ & $\rho^4$ & $\rho^4\ln\rho$ & $\rho^4\ln^2\rho$ & $\rho^4\ln^3\rho$ \\ 
\hline
$X_A$ &
3.5588 &
0 &
 $- 145.23$ & $- 105$ & $- 36$ &
 $- 332.91$ & 0 &
1807.0 & $- 130.61$ & 138.48 & $- 144$ \\
$X_N$ &
 $- 9.0464$ &
0 &
 $- 125.74$ & $- 61.667$ & 22 &
 $- 182.54$ & $- 407.94$ &
 $- 525.17$ & $- 298.36$ & $- 135.65$ & 88 \\
$X_C$ &
3.2203 &
 $- 12.337$ &
100.97 & 17.333 & $- 8$ &
155.99 & 210.55 &
112.72 & 343.28 & 44 & $- 32$ \\
$X_L$ &
3.2203 &
0 &
26.174 & 17.333 & $- 8$ &
23.121 & 105.28 &
41.907 & $- 13.638$ & 52 & $- 32$ \\
$X_H$ &
 $- 0.036433$ &
0 &
 $- 0.30328$ & 0 & 0 &
0 & 0 &
 $- 8.8409$ & $- 8.2856$ & 0 & 0 \\
$L^{(1)}_2$ &
 $- 8.1819$ &
 $- 3.4544$ &
 $- 320.89$ & $- 185.83$ & 1.25 &
 $- 581.80$ & $- 640.43$ &
 $- 257.71$ & $- 1141.8$ & $- 245.37$ & 13.111 \\
$L^{(2)}_2$ &
 $- 3.3806$ &
 $- 1.5353$ &
 $- 160.06$ & $- 90.557$ & 0.6 &
 $- 290.22$ & $- 320.21$ &
 $- 820.05$ & $- 951.38$ & $- 215.13$ & 10.444 \\
$H^{(1)}_2$ &
 $- 6.1150$ &
 $- 1.8094$ &
 $- 28.786$ & $- 46.563$ & 0.41667 &
4.0514 & 0 &
1648.7 & 914.98 & 276.28 & $- 2.6667$ \\
$H^{(2)}_2$ &
 $- 1.7910$ &
 $- 0.3838$ & 
 $- 3.8498$ & $- 6.9712$ & 0.066667 &
0.90167 & 0 &
 $- 645.37$ & $- 317.85$ & $- 26.783$ & 0 \\
$U_C$ &
3.2203 &
 $- 12.337$ &
47.319 & 0 & 0 &
119.90 & 105.28 &
 $- 104.95$ & 120.05 & $- 8$ & 0
\end{tabular}
\begin{tabular}{l  r  r  r  r  r  r  r  r }
& $\rho^5$ & $\rho^5\ln\rho$ & $\rho^6$ & $\rho^6\ln\rho$ & $\rho^6\ln^2\rho$ & $\rho^6\ln^3\rho$ & $\rho^7$ & $\rho^7\ln\rho$ \\ 
\hline
$X_A$ &
 $- 617.86$ & 1579.1 & 
 $- 610.71$ & 91.319 & 95.033 & 0 &
 $- 39.188$ & 301.79 \\
$X_N$ &
263.65 & $- 981.70$ &
540.82 & $- 121.07$ & $- 12.203$ & 0 &
 $- 52.784$ & $- 158.92$ \\
$X_C$ &
 $- 188.62$ & 164.49 &
 $- 235.84$ & 87.674 & $- 7.111$ & 0 &
144.75 & 177.65 \\
$X_L$ &
 $- 120.76$ & 105.28 &
22.976 & $- 57.926$ & 24 & 0 &
18.799 & 0 \\
$X_H$ &
0 & 0 &
7.5624 & $- 12.813$ & 3.2 & 0 &
0 & 0 \\
$L^{(1)}_2$ &
 $- 392.83$ & $- 1291.7$ &
1458.5 & $- 757.80$ & 127.03 & $- 8.7037$ &
141.85 & $- 48.398$ \\
$L^{(2)}_2$ &
 $- 286.40$ & $- 1359.2$ &
1080.5 & $- 1028.9$ & 13.504 & $- 16.684$ &
425.16 & $- 211.77$ \\
$H^{(1)}_2$ &
359.98 & 1571.0 &
 $- 1727.4$ & 1184.0 & $- 191.14$ & 29.728 &
 $- 346.40$ & 110.98 \\
$H^{(2)}_2$ &
 $- 358.33$ & $- 844.98$ &
78.867 & $- 1134.4$ & $- 176.54$ & $- 34.547$ &
420.77 & $- 560.13$ \\
$U_C$ &
 $- 73.693$ & 0 &
34.956 & 1.6 & $- 7.1111$ & 0 &
 $- 33.839$ & 0
\end{tabular}
\end{ruledtabular}
\caption{\label{tab:results} 
Coefficients of $\rho^i \ln^j \rho$ in expansions of the components of
  $X_2$, Eq.~(\protect\ref{eqn:X2}), and of the
  moments $\langle \hat{E}_{l,h}^{1,2} \rangle$.  The last line shows
  the analogue of $X_C$ for the decay $b\to u\ell\bar{\nu}$.  }
\vspace{-0.5cm}  
\end{table*} 
\endgroup

Our results can be tested by comparing logarithms of the mass ratio
with the values predicted by the renormalization group analysis
\cite{Bauer:1996ma}.  That study summed up some of the logarithmic
effects to all orders in the coupling constant.  When those results
are expanded in $\alpha_s$, three terms can be tested in order
$\alpha_s^2$: $\rho^2 \ln^2 \rho$, $\rho^3 \ln \rho$, and $\rho^4
\ln^3 \rho$.  They are related to the presence in the tree-level
decay width (\ref{eq:tree}) of terms $\rho^2$ and $\rho^4 \ln\rho$, and
the absence of $\rho^3$ there.

Ref.~\cite{Bauer:1996ma} traced the origin of those terms to operators
that can be constructed from the $b$ and $c$ quark fields and
determined anomalous dimensions of those operators.  The logs in
terms $\rho^2$ were found to arise from the running of the $c$-quark
mass, and those in $\rho^4$ -- from a complicated mixing of a variety
of dimension seven operators.  Our results fully agree with that
analysis.

However, terms $\rho^3\ln \rho$ turn out to work differently. In
\cite{Bauer:1996ma} they were attributed to the running of
$m_c$, with the resulting coefficient that disagrees with our result.
We find that those terms originate from the four-quark operator
$\overline h_b \Gamma_\mu  c \ \overline c \Gamma^\mu  h_b$.
Here $h_b$ denotes the static field describing a slow quark $b$. As
discussed in Ref.~\cite{Bauer:1996ma}, this operator gives rise to
terms $\rho^3$ in the tree-level decay in the case of a {\em vector}
coupling ($\Gamma_\mu = \gamma_\mu$), but does not contribute at the
tree-level in the chiral case ($\Gamma_\mu = \gamma_\mu(1-\gamma_5)$).
However, we find that at $\order{\alpha_s}$ it has a finite matrix
element as shown in Fig.~\ref{fig:regions}(d), responsible for the cubic
mass term in the decay width (\ref{eq:X1}).

At the next loop level, the effects of the coupling constant running
and of the anomalous dimension of that operator generate terms
$\alpha_s^2 \rho^3 \ln \rho$.  Including charm, $N_L+1$
quarks contribute to the coupling running.  Denoting $\beta_0 \equiv
{11\over 3}C_A - {4\over 3}T_R(N_L+1)$, one finds
\begin{equation}
-8\beta_0 + {32\over 3}T_R - 12 C_A = {32\over 3}T_RN_L +{64\over
 3}T_R - {124\over 3}C_A,
\end{equation}
in agreement with the coefficients of $\pi^2\rho^3 \ln \rho$ in
Eq.~(\ref{eqn:xn}).

The linear term $-5\pi^2\rho/4$ in Eq.~(\ref{eqn:xn}) is noteworthy.
As will now be explained, it arises because the on-shell (pole)
definition of the $b$-quark mass has been used.  Although not suitable
for phenomenology, it has been adopted for the ease of comparisons
with Ref.~\cite{Kirill2008} and simplicity of presentation.  The
linear term $m_q/m_b$ arises from a $q$-quark ($m_q \ll m_b$) loop
inserted in the gluon propagator in Fig.~\ref{fig:regions}(d).  Thus
it does not depend on the final-state quark mass and equally affects
decays $b\to c\ell \bar{\nu}$ and $b\to u\ell \bar{\nu}$.  Arising from the gluons
with momentum $\order{m_q}$ this effect becomes a problem for the
perturbative analysis when $q$ is light, $m_q \leq \Lambda_{\rm QCD}$.
This illustrates how a linear $\Lambda_{\rm QCD}/m_b$ correction
appears even in the total $b$-quark width when the pole mass is used.
If a short-distance mass definition is used, such as the
$\overline{\mbox{MS}}$ mass \cite{gbgs90}, such terms are absorbed
into the lowest-order decay width and are absent in higher-orders of
the $\alpha_s$ expansion
\cite{Chay:1990da,Bigi:1992su,Shifman:1986mx}.  Note the
factor five in the coefficient of the linear term, related to the
fifth power of $m_b$ in the width formula.

Where the linear correction really counts is the muon decay, whose
width is traditionally expressed using the muon pole mass $m_\mu$.
The previous study of the muon decay \cite{vanRitbergen:1999fi}
neglected the electron mass $m_e$. Its effect on the decay rate was
assumed to arise from terms $ \left( \alpha\over \pi\right)^2
\left(m_e\over m_\mu\right)^2 \ln^2\left(m_e\over m_\mu\right)^2\simeq
1.5\cdot 10^{-8}$.  The theoretical error was estimated by taking the
coefficient of this term to be 24.  Due to the overlooked linear
correction the electron mass effect turns out to be even larger,
affecting the determination of the Fermi constant from the anticipated
new measurement.

The Fermi constant is determined from the measured muon lifetime using
the relation \cite{vanRitbergen:1999fi}
\begin{equation}
{1\over \tau_\mu } 
  = {G_F^2 m_\mu^5 \over 192 \pi^3} \left( 1 + \Delta q\right),
\end{equation}
where $\Delta q$ describes effects of the finite electron mass and
radiative corrections in the limit of the four-fermion contact
interaction ($m_\mu \ll M_W$).  The latter have been known exactly in
the first order in $\alpha$, and in the limit of zero $m_e$ in order
$\alpha^2$.  Present result extends the knowledge of the $\alpha^2$
correction to the case of the finite $m_e$.  The extra shift is
$\Delta q (m_e) \simeq -0.43 \cdot 10^{-6}$, comparable with the
expected precision of the MuLan result.

Even though the coefficient of the $ \left( \alpha\over \pi\right)^2
\left(m_e\over m_\mu\right)^2
\ln^2\left(m_e\over m_\mu\right)^2$ term is only $-11$, less than half
the value 
taken in \cite{vanRitbergen:1999fi} for the purpose of an error
estimate, the total electron mass effect, dominated by the linear term
$m_e/m_\mu$,  is larger than expected.

For the decay $b\to c \ell \bar{\nu}$, we confirm numerical results
of \cite{Kirill2008}.  The flexible numerical method
developed in that reference enables one to impose phase space cuts.
Our approach is complementary.  It provides analytical results and
gives insight into the small-mass region, important for the muon
decay.  It also facilitates changes of the scheme used for the heavy
quark masses.  Finally, it reveals the structure of logarithms and
highlights the relative importance of various operators, improving on
the analysis of \cite{Bauer:1996ma}.  But is it possible to combine
the analytical approach with cuts on the lepton energy?  

According to \cite{Kirill2008},
effects of cuts for the {\em lepton energy} moments can
be modeled with the tree-level distribution \cite{cj94}.  For the cut
$E_l > E_{\rm cut} = 1$~GeV one finds $L_{2{,\rm
cut}}^{(n)}/L_{2}^{(n)} \approx L_{1{,\rm cut}}^{(n)}/L_{1}^{(n)}
\approx L_{0{,\rm cut}}^{(n)}/L_{0}^{(n)}$.  For the rate
$L_{2{,\rm cut}}^{(0)}$ the error is only $-4$\%, and less than $-2$\%
for $L_{2{,\rm cut}}^{(1,2)}$, indicating smallness of the hard-gluon
radiation.

The relative impact of cuts on the {\em hadron energy} moments
$\langle\hat{E}_h^{n}\rangle$ depends on the order $n$ of the moment
only very weakly \cite{Kirill2008}.  To within 1\%, $H_{2{,\rm
cut}}^{(1)}/H_{2}^{(1)}$ and $H_{2{,\rm cut}}^{(2)}/H_{2}^{(2)}$ 
both equal $L_{2{,\rm cut}}^{(0)}/L_{2}^{(0)}$ (and significantly
exceed $H_{0{,\rm cut}}^{(1,2)}/H_{0}^{(1,2)}$).  First-order
corrections behave similarly: $H_{1{,\rm cut}}^{(1)}/H_{1}^{(1)}
\approx H_{1{,\rm cut}}^{(2)}/H_{1}^{(2)} \approx L_{1{,\rm
cut}}^{(0)}/L_{1}^{(0)}$ within 0.1\%.
Thus the effect of cuts $E_{\rm cut} \lesssim 1$~GeV on 
QCD corrections to
$\langle\hat{E}_h^{n}\rangle$ can be
approximated by that on the rate.

Our result for the correction to the width exceeds the earlier
estimate based on an interpolation of special kinematic results
\cite{Czarnecki:1998kt}.  The full mass dependence being now known, it
is clear that the correction varies strongly as a function of the mass
ratio and is close to vanishing near the physical value.  This
indicates cancellations that were not taken into account in 
\cite{Czarnecki:1998kt}.  In particular, the non-BLM correction turns
out to be about $1.7(\alpha_s/\pi)^2$ \cite{Kirill2008}.  While the
full phenomenological analysis requires a fit of the rate and moments,
we expect this correction to decrease the value of $|V_{cb}|$
determined from inclusive $b$ decays by about one percent,
bringing it slightly closer to the exclusive-decay result.

We thank K. Melnikov for sharing his results prior to publication and
many helpful discussions. We also thank V.~A.~Smirnov for providing
unpublished results on master integrals.  We are grateful to
I.~R.~Blokland, M.~Yu.~Kalmykov, A.~V.~Kotikov, A.~A.~Penin,
J.~Piclum, A.~Vainshtein for advice and to I. W. Stewart for bringing
Ref.~\cite{Bauer:1996ma} to our attention. This research was supported
by the Science and Engineering Research Canada.

\end{document}